\documentclass[final,1p,times]{elsarticle} % do not change
\usepackage{graphicx} % do not change
\usepackage{amssymb} % do not change
\usepackage{amsthm} % do not change
\usepackage{lineno} % do not change

\journal{Nuclear Physics A} % do not change
\begin{document} % do not change

\begin{frontmatter} % do not change

%% QM09Author: please enter your  
%% Title, author and address info here; please do not use footnotes

% Your Title - please modify
\title{Jet quenching and the gravity dual}

% Principle author, and co-authors - please modify
\author{Mohammed Mia and Charles Gale}

% Address - please modify
% note that if you have authors from several institutions, we recommend
% labelling these [a], [b], [c] etc.
\address{Department of Physics, McGill University, 3600 rue University
Montreal, QC, Canada H3A 2T8}
\begin{abstract} % do not change
%% Text of abstract goes here - please modify
We compute the momentum broadening of an energetic parton as
it moves through strongly interacting matter, using the gravity dual of a non conformal gauge 
theory with matter in the fundamental representation. Our approach defines a family
of theories, all of which have logarithmic running of the couplings
in the far IR but become almost conformal in the far UV. 
For a given set of degrees of freedom, we obtain average transverse momentum
square of the parton as a function of
temperature and compare with results obtained from the AdS/CFT correspondence.
\end{abstract} % do not change
\end{frontmatter} % do not change

%% QM09: we keep linenumbers at least for initial version
%\linenumbers % do not change

%% start of main text - please modify. Below is a sub-set (single section) 
%% of an earlier proceedings that show how one can handle references 
%% and figures etc.
%%\section{}\label{}

\section{Introduction}
Understanding the dynamics of the fluid created in the early 
stages of a heavy ion collision is a challenge, as empirical evidence at RHIC indicates that interactions are strong.  However, duality stipulates that the Hilbert space of strongly coupled gauge theory is
contained in the Hilbert space of weakly coupled gravity. Thus the analysis of the weakly coupled dual geometry of strongly coupled thermal field theory may be relevant for the physics of the quark-gluon plasma. In this note, we
attempt to construct the dual gravity of a strongly coupled thermal gauge theory
which has logarithmic running of the coupling in the far IR but  becomes
almost conformal in the far UV. The gauge theory has matter in the fundamental
representation, a small number of flavors but a large number of colors and thus
shares common features with large-$N$ QCD. Using this geometry, we compute
the transverse momentum broadening of a fast moving parton as it travels through 
the QCD-like plasma. As distinct geometries describe gauge theories with different
degrees of freedom, our method can be used in principle to study various phases of a strongly interacting medium.     
\begin{figure}[ht]
\begin{center}
\hspace*{-1cm}
\includegraphics[width=0.45\textwidth]{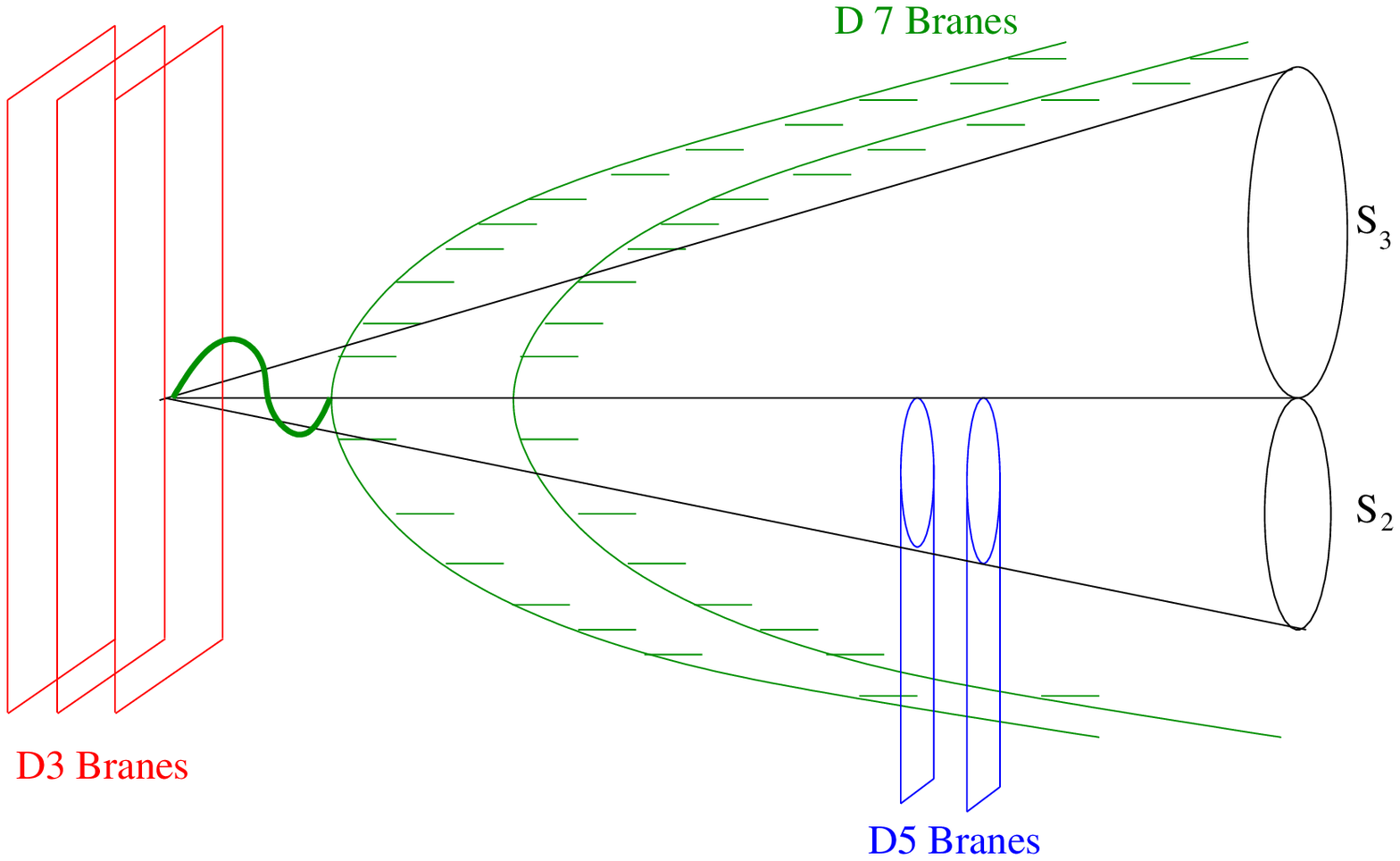}
\parbox{6cm}{\vspace*{-4cm} \includegraphics[width=0.45\textwidth]{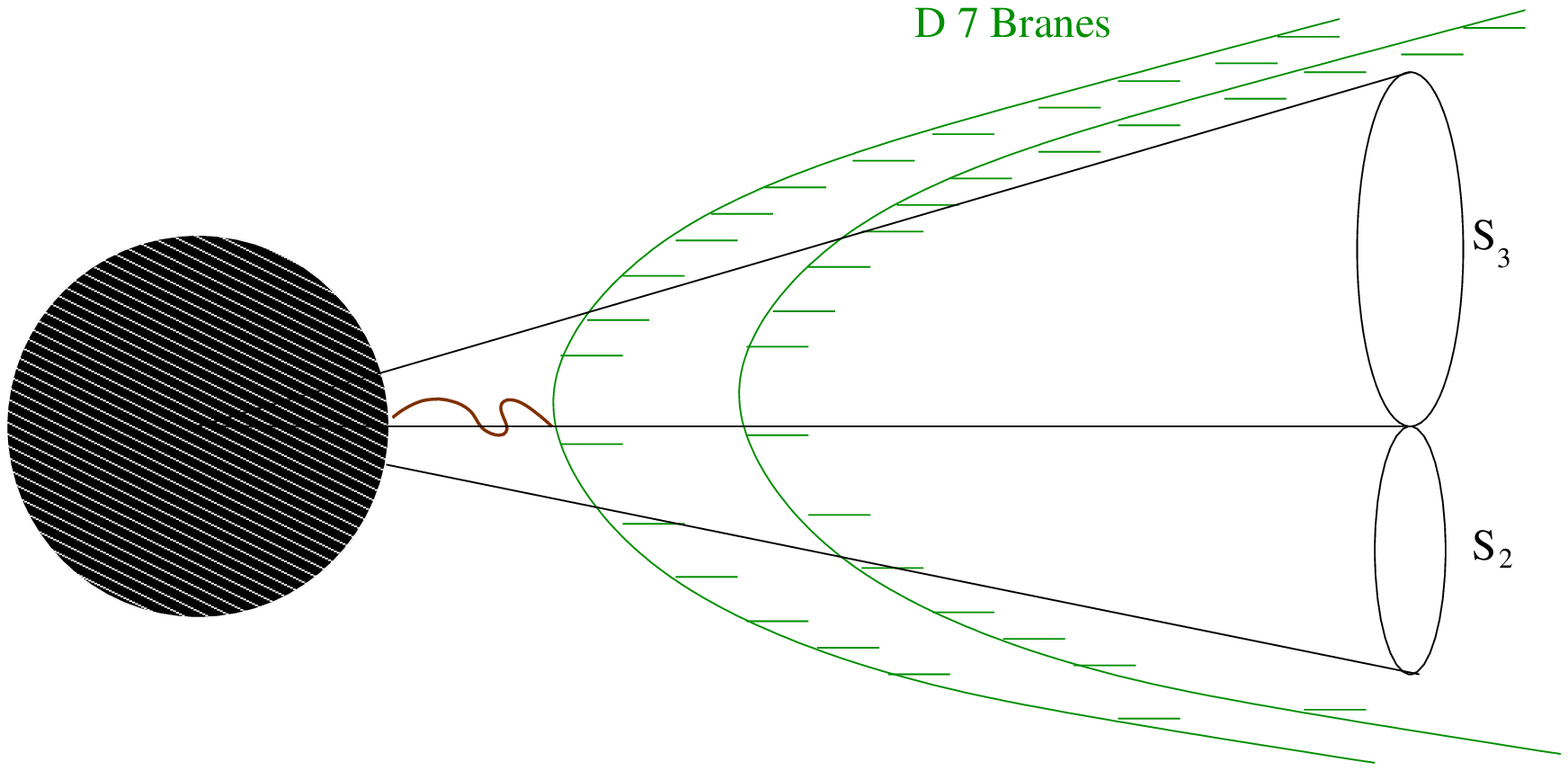}}
\end{center}
\caption{Left figure is the brane configuration in our approach while the figure on the right
represents the dual geometry of thermal field theory resulting from the brane
excitations.}
\label{geometry}
\end{figure}
\section{Construction of the geometry}
Consider the following set up: place $N$ D3 branes at the tip of a six-dimensional conifold with base $T^{1, 1}=S^2\times S^3$ with the D3 branes world volume
extended in the four Minkowski directions. Now wrap $M$ D5 branes around the
$S^2$ of the base with the remaining world volume extending in the Minkowski
directions. Finally if one embeds $N_f$ D7 branes, the resulting gauge theory has
$SU(N+M)\times SU(N)$ color symmetry with $SU(N_f)\times SU(N_f)$ flavor
symmetry and the gauge couplings $\tilde{g}_1, \tilde{g}_2$ run logarithmically
with scale $\Lambda$,    
$\partial/\partial \ln \Lambda \left[4\pi^2 / \tilde{g}_1^2 +
4\pi^2 / \tilde{g}_2^2\right] = 
- 3N_f / 4$, 
$\partial / \partial \ln \Lambda\left[4\pi^2 / \tilde{g}_1^2 -
4\pi^2 / \tilde{g}_2^2 \right] = 
3M\left[1 + \left(3g_s N_f / 2\pi\right) \ln \Lambda\right]$, where $g_s$ is the string coupling constant. 
The gauge theory described above is at zero temperature. The dual of
the finite temperature gauge theory is a ten-dimensional geometry with a black
hole horizon $r_h$ and after dimensional reduction the five-dimensional geometry
is a modified AdS space with the metric \cite{FEP}
\begin{eqnarray}\label{bhmet}
ds^2_5 &=& {1\over \sqrt{h}}
\left(-g_1dt^2+d\overrightarrow{x}^2\right)+\frac{\sqrt{h}}{g_2}dr^2={1\over \sqrt{h}}\left(g_{ij}dx^idx^j\right)+\frac{\sqrt{h}}{g_2}dr^2
\end{eqnarray}
where $i, j=0, 1, 2, 3$, $h=L^4 / r^4 \left(1+A\, {\ln }(r/r_0)+B\,   \ln^2(r/r_0)\right)$, $r_0$ is a constant determined 
by the location of
the branes, $g_1, g_2$ are functions of $r$ and 
$A, B$ are ${\cal O}(g_sM^2/N, g_s^2N_f) $ constants \cite{FEP}. We can use the 
supergravity action describing this geometry to find the gauge theory partition
function \cite{WM} ${\cal Z}_{\rm Gauge}[\phi_0]\equiv {\rm exp}(S^{\rm total}_{\rm string}[\phi_0])
\simeq{\rm exp}(S^{\rm total}_{\rm SUGRA}[\phi_0])$.
 The gauge theory has couplings which run logarithmically from UV to IR and the number of effective degrees
of freedom blows up in the UV. It turns out that by placing further D branes in a specific configuration \cite{FEP,BVYA}, the running of the couplings
can be modified in the UV so that $\tilde{g}_i\sim \frac{c^n}{\Lambda^n}$ for
large $\Lambda$, where $c^n$ are some constants. The dual geometry of this modified gauge theory will have the 
metric of the form (\ref{bhmet}), with a 
warp factor $h\sim\frac{L^4}{r^4}\left(\frac{\tilde{A}_n}{r^n}\right)$ for large $r$, $\tilde{A}_n$ being constants. 
The warp factor has to be continuous across some intermediate value $r=r_c$ and demanding that the energy momentum tensor is
continuous forces $dh/dr,d^2h/dr^2$ to be continuous across  $r=r_c$. Using the continuity of the warp factor and
its derivatives at $r=r_c$, one can express $\tilde{A}_0, \tilde{A}_1, \tilde{A}_2$ as a functions of $A, B$. We will set
$\tilde{A}_n=0$ for $n>2$ which is sufficient for continuity. Thus the dual five-dimensional geometry of our desired gauge 
theory will be of the form  (\ref{bhmet}) with  
\begin{eqnarray} \label{KS16}
&&h=\frac{L^4}{r^4}\left(1+A\, \ln (r/r_0)+B\, \ln^2 (r/r_0) \right), ~~{\rm for}~~  r<r_c\nonumber\\
&&h=\frac{L^4}{r^4}\left(\tilde{A}_0+\frac{\tilde{A}_1}{r}+\frac{\tilde{A}_2}{r^{2}}\right), ~~{\rm for}~~ r>r_c
\end{eqnarray}

\section{Jet quenching from geometry}
\begin{figure}[ht]
\begin{center}
\includegraphics[width=\textwidth]{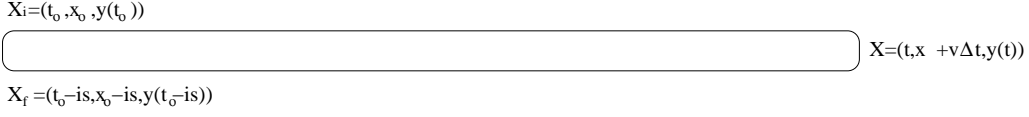}
\caption[]{The contour C with $X$ denoting the coordinates of a point on the contour. The upper line with real time coordinate is the
worldline of the heavy parton while the lower line has complex time coordinate.
Fields evaluated on the upper line are the type `1' fields while those evaluated
at lower line are type `2' fields of thermal field theory. Here $s\rightarrow 0$ is real.}
\label{charm_xsec}
\end{center}
\end{figure}
Using the Wigner distribution function $f(X, k_\perp)$ as in QCD kinetic theory \cite{KT}, we will compute the average transverse momentum 
square $<p^2_\perp>=\int d^3x\int\frac{d^2k_\perp}{(2\pi)^2}k^2_\perp f(X, k_\perp)$ of a heavy fast moving parton we call the jet.
It turns out that one can express the distribution function and subsequently $<p^2_\perp>$ solely in terms of functional derivatives of Wilson 
loops \cite{DTCS} and if we assume that the initial transverse momentum distribution is narrow, then $<p^2_\perp>=2\kappa_T {\cal T}$ with ${\cal T}$ some large time
interval, and $\kappa_T$ is the diffusion coefficient. As shown in \cite{DTCS}, one can write $\kappa_T=\lim_{\omega\rightarrow 0}\frac{1}{4}
\int dt \;e^{i\omega t}\;(iG_{11}(t, 0)+iG_{22}(t, 0)+iG_{12}(t, 0)+iG_{21}(t, 0))$ where $G_{ij}\sim \frac{\delta^2 W_C}{\delta y_i\delta y_j}$ are Green's functions
with  $W_C$ being the Wilson loop. Here $C$ is the contour in Fig. 2.

We will now compute the Wilson loop at strong coupling by using holography, that is we  identify $<{\rm tr}\rho^0
W_C>=e^{iS_{\rm NG}}$, 
$\rho^0$ is the density matrix \cite{DTCS}, $S_{\rm NG}$ being the Nambu-Goto action, with the boundary of the string worldsheet being  the curve $C$. We consider a heavy parton moving in the $x$ direction with velocity
$v$ and transverse displacement in the $y$ direction, where $t, x, y, z$ are the four Minkowski coordinates. If $X^\mu:(\sigma, \tau)\rightarrow
(t, r, x, y, z)$ is a mapping from string worldsheet to five-dimensional geometry given by (\ref{bhmet}), then with parametization
$\sigma=r, t=\tau$, we have $x(t, r)=vt+\bar{x}(r)$. Using the solution $\bar{x}$ as in \cite{FEP} with
 $g_1=g_2=g$, and changing coordinate $\zeta=1/r$, 
we find the worldsheet metric to be off diagonal with components $f_{tt},f_{t\zeta},f_{\zeta\zeta}$ being
function of $\zeta,t$ and transverse displacement $y$.
  This worldsheet metric can be
diagonalized at zeroth order in $y$ with the reparametization
$\hat{t}= 1/\sqrt{\gamma} \left( t + F(\zeta) \right), \hat{\zeta}=\sqrt{\gamma}\zeta$ and  
the resulting worldsheet metric is
\begin{eqnarray}
f_{\hat{t}\hat{t}}&=&h(\zeta)^{-1/2} \left( - g(\hat{\zeta}) + \left( d\hat{y}/d\hat{t}\right)^2 \right)/
\gamma ,~~~f_{\hat{t}\hat{\zeta}}={\cal O}(y^2);\nonumber\\
f_{\hat{\zeta}\hat{\zeta}}&=&\sqrt{h(\zeta)}\left(\gamma \zeta^4 g(\hat{\zeta}\right)^{-1} 
 \left(1+{\cal
O}(g_s\tilde{N})\right)+\left(d\hat{y}/d\hat{\zeta}\right)^2 / \left(\gamma \sqrt{h(\zeta)}\right)
\end{eqnarray} 

Here $\gamma=1/\sqrt{1-v^2}$, $\tilde{N}={\cal O}(M^2/N)+{\cal O}(g_sN_fM^2/N)$,  
$g(\hat{\zeta})=1- \hat{\zeta}^4/\zeta_{h}^4$, $\zeta_h=1/r_h$ with $r_h$ the blackhole horizon of spacetime,
 $\hat{y}=\sqrt{\gamma}y$ and the precise form of $F(\zeta)$ is determined
by the off diagonal worldsheet metric \cite{qhat1}. 
With this reparametization, the
Nambu-Goto action, to quadratic order in $\hat{y}$, becomes 
\begin{eqnarray}
S_{NG}^{[2]}=\frac{1}{2\pi \alpha'}\int d\hat{\zeta}d\hat{t}
\frac{1-\frac{{\cal A}(\hat{\zeta})}{2}}{2\hat{\zeta^2}}\left(-\frac{\dot{\hat{y}}^2(1+{\cal A}(\hat{\zeta}))}{g(\hat{\zeta})}+\frac{\hat{\zeta}^4g(\hat{\zeta})\hat{y}'^2}{\gamma^2
h(\zeta)}\right)
\end{eqnarray}  
where $\dot{\hat{y}}=d\hat{y}/d\hat{t}, \hat{y}'=d\hat{y}/d\hat{\zeta}$. If we denote $\hat{y}(\hat{t}, \hat{\zeta})=\int
d\hat{\omega} \;e^{i\hat{\omega} \hat{t}}\; \hat{y}(\hat{\omega})\hat{Y}(\hat{\zeta})$, then the equation of motion can be written as  
\begin{eqnarray} \label{EOM}
\hat{Y}''+\frac{B'}{B}\hat{Y}'-\frac{D}{B}\hat{Y}=0, ~~B=\frac{\hat{\zeta}^2g(\hat{\zeta})}{2\gamma^2h(\zeta)}\left(1-\frac{{\cal A(\hat{\zeta})}}{2}\right),
~~D=-\frac{\hat{\omega}^2(1+\frac{{\cal A}(\hat{\zeta})}{2})}{2\hat{\zeta}^2g(\hat{\zeta})}
\end{eqnarray} 
where prime denotes derivative with respect to $\hat{\zeta}$, and ${\cal A}(\hat{\zeta})$ is of ${\cal O}(g_s\tilde{N})$. The solution to
(\ref{EOM}) takes the form $\hat{Y}=g(\hat{\zeta})^{-i\frac{\hat{\omega}}{4\pi T}}{\cal F}$, ${\cal F}=(1+i\frac{\hat{\omega}}{4\pi T}H)$ where
$T=\frac{\zeta_h}{\pi\sqrt{h(\zeta_h)}}$ is the temperature and we have written ${\cal F}$ only up to linear order in $\hat{\omega}$. $H$ can be
expanded as a power series in $\hat{\zeta}$ for small $\hat{\zeta}$, $H=a_i\hat{\zeta}^i$. For large $\hat{\zeta}>\hat{\zeta}_c=\sqrt{\gamma}/r_c$, 
$H=\sum_{i, j=0}^{\infty}\tilde{a}_{ij}\hat{\zeta}^i{\rm log}^j(\hat{\zeta})$ . Note that $\tilde{a}_{ij}={\cal O}(g_s\tilde{N})$ for $j>0$, 
whereas $\tilde{a}_{i0}$ is zeroth order in
$g_s\tilde{N}$ for every $i$. At zeroth order in $g_s\tilde{N}$, $H$ can be written in a closed form
\begin{eqnarray}
H^{(0)}=2{\rm tan}^{-1}\left(\frac{\hat{\zeta}}{\zeta}_h\right)-2 \ln \left(1+\frac{\hat{\zeta}}{\zeta_h}\right)- \ln \left(1+\frac{\hat{\zeta}^2}{\zeta_h^2}\right)
\end{eqnarray} 
for all $\hat{\zeta}$, where we have imposed boundary condition $H^{(0)} (0)=0$. For linear order in $g_s\tilde{N}$ and small
$\hat{\zeta}$, $H^{(1)}=\sum_{i=3} b_i\hat{\zeta}^i$ with $b_i$ of ${\cal O}(g_s\tilde{N})$, 
equation (\ref{EOM}) demands $b_1=b_2=0$ and we have set $b_0=0$ complying with the boundary condition $H^{(1)}(0)=0$. 
As we have a second order differential equation, one of the $b_i$'s is still undetermined. Now for large
$\hat{\zeta}$, imposing boundary condition $H(\zeta_h)=0$ fixes one of the $\tilde{a}_{ij}$'s and the equation of motion leaves 
one of the $\tilde{a}_{ij}$'s undetermined. But $H, H'$ and $H''$ have to be continuous across 
$\hat{\zeta}_c$. Once we set $H$ and $H'$ to be continuous
across  $\hat{\zeta}_c$, the equation of motion guarantees that $H''$ is also continuous. This means we have two equations making $H, H'$
continuous and these fix the two unknowns (one of $\tilde{a}_{ij}$ and one of $b_i$). Hence we have the solution $H$  
for all $\hat{\zeta}$  with all coefficients determined \cite{qhat1}. Now using the solution for $\hat{Y}$ and taking appropriate linear 
combinations to build the type `1' and `2' fields
$y_1, y_2$ as in \cite{DTCS}, we can write the boundary action after integrating  the Nambu-Goto action and the result is 
equation (3.51) of 
 \cite{DTCS} but $\hat{Y}$ is replaced with our solution, $\hat{\omega}, \omega$ replaced by $\hat{\omega}/\pi T, \omega/\pi T$ and
 $R=1$.
 Finally from the boundary action we can obtain the Green's function $G_{ij}$ and the result
 for the diffusion coefficient is
\begin{eqnarray} \label{NG1}
\kappa_T=\sqrt{\gamma g_sN_{\rm eff}}\pi T^3 (1+{\cal B})
\end{eqnarray}            
where $\cal{B}$ is of ${\cal O}(g_s\tilde{N})$ \cite{qhat1} and $N_{\rm eff}$ is the number of effective degrees of freedom for the boundary 
gauge theory
\begin{eqnarray} \label{Neff}
N_{\rm eff}=N\left(1+\frac{27g_s^2M^2N_f}{32\pi^2 N}-\frac{3g_sM^2}{4\pi N}+\left[\frac{3g_sM^2}{4\pi
N}-\frac{9g_s^2M^2N_f}{16\pi^2 N}\right]{\rm log}\left(\frac{r_c}{r_0}\right)
+\frac{9g_s^2M^2N_f}{8\pi^2N}{\rm log}^2 \left(\frac{r_c}{r_0}\right)\right)
\end{eqnarray} 
where $r_c$ is as in (\ref{KS16}), indicating where the warp factor changes from a logarithm to a power series. The value of
$r_c$ will of course be determined by the precise location of the branes. Thus for different configuration of branes in
ten dimensions, we will end up with different effective degrees of freedom for the gauge theory in four-dimensional
Minkowski space. Note for duality to hold, $N$ must be quite large, making $N_{\rm eff}$ quite large. For $N_f=M=0$, we get
back the value of $\kappa_T$ as computed in \cite{DTCS}. However for a non conformal field theory with fundamental matter
- which is more relevant for QCD - $M\neq 0, N_f \neq 0$, and our analysis thus generates a correction to the AdS/CFT result. 

\section*{Acknowledgments} % please check/modify, comment out or delete if not needed
We would like to thank K. Dasgupta and S. Jeon for much help and discussions. M. M. is happy to acknowledge financial support from the organizers of Quark Matter 2009. This work is funded in part by the Natural Sciences and Engineering Research Council of Canada.

 % do not change 
\end{document}